\documentclass[sigconf]{acmart}
\AtBeginDocument{%
  }

\usepackage{tabularx}
\usepackage{multirow}
\usepackage{multicol}
\usepackage{url}
\usepackage{booktabs}
\usepackage{xcolor}
\usepackage{colortbl}
\usepackage{algorithm}
\usepackage{algorithmic}
\usepackage{tcolorbox}
\usepackage{pifont}
\usepackage{relsize}
\usepackage[inline]{enumitem}
\usepackage{array, xcolor, colortbl}
\usepackage{booktabs}  
\usepackage{tcolorbox} 
\usepackage{afterpage}

\newcommand{\ours}{\textsc{RankFlow}\,}
\definecolor{lightgreen}{rgb}{0.88, 1, 0.88}
\copyrightyear{2025}
\acmYear{2025}
\setcopyright{cc}
\setcctype{by}
\acmConference[WWW Companion '25]{Companion Proceedings of the ACM Web Conference 2025}{April 28-May 2, 2025}{Sydney, NSW, Australia}
\acmBooktitle{Companion Proceedings of the ACM Web Conference 2025 (WWW Companion '25), April 28-May 2, 2025, Sydney, NSW, Australia}
\acmDOI{10.1145/3701716.3717575}
\acmISBN{979-8-4007-1331-6/2025/04}

\begin{document}

\title{\ours: A Multi-Role Collaborative Reranking Workflow Utilizing Large Language Models}
\author{Can Jin}
\authornote{Both authors contributed equally to this research.}
\email{can.jin@rutgers.edu}
\affiliation{%
  \institution{Rutgers University}
  \city{Piscataway}
  \state{New Jersey}
  \country{USA}
}

\author{Hongwu Peng}
\authornotemark[1]
\email{hongwu.peng@uconn.edu}
\affiliation{%
  \institution{University of Connecticut}
  \city{Storrs}
  \state{Connecticut}
  \country{USA}
}

\author{Anxiang Zhang}
\email{adamzhang1679@gmail.com}
\affiliation{%
  \institution{Independent}
  \city{Mountain View}
  \state{California}
  \country{USA}
}

\author{Nuo Chen}
\email{pleviumtan@toki.waseda.jp}
\affiliation{%
  \institution{Waseda University}
  \city{Shinjuku City}
  \state{Tokyo}
  \country{Japan}
}

\author{Jiahui Zhao}
\email{jiahui.zhao@uconn.edu}
\affiliation{%
  \institution{University of Connecticut}
  \city{Storrs}
  \state{Connecticut}
  \country{USA}
}

\author{Xi Xie}
\email{xi.xie@uconn.edu}
\affiliation{%
  \institution{University of Connecticut}
  \city{Storrs}
  \state{Connecticut}
  \country{USA}
}

\author{Kuangzheng Li}
\email{likuangzheng0633@gmail.com}
\affiliation{%
  \institution{Independent}
  \city{Storrs}
  \state{Connecticut}
  \country{USA}
}

\author{Shuya Feng}
\email{shuya.feng@uconn.edu}
\affiliation{%
  \institution{University of Connecticut}
  \city{Storrs}
  \state{Connecticut}
  \country{USA}
}

\author{Kai Zhong}
\email{kaizhong@gmail.com}
\affiliation{%
  \institution{Independent}
  \city{Palo Alto}
  \state{California}
  \country{USA}
}

\author{Caiwen Ding}
\email{caiwen.ding@uconn.edu}
\affiliation{%
  \institution{University of Connecticut}
  \city{Storrs}
  \state{Connecticut}
  \country{USA}
}

\author{Dimitris N. Metaxas}
\email{dnm@cs.rutgers.edu}
\affiliation{%
  \institution{Rutgers University}
  \city{Piscataway}
  \state{New Jersey}
  \country{USA}
}
\renewcommand{\shortauthors}{Can Jin et al.}

\begin{abstract}
In an Information Retrieval (IR) system, reranking plays a pivotal role by ordering candidate passages based on their relevance to a specific query. This process necessitates a fine-grained understanding of the variations among passages associated with the query. However, existing zero-shot reranking methods often rely on the original queries and passages from the dataset without modification, which can sometimes be vague, ambiguous, or redundant. In this work, we introduce \ours, a multi-role reranking framework designed to address these limitations and enhance reranking performance. \ours decomposes the reranking task into four components by leveraging large language models to perform four distinct roles: the query \textbf{Rewriter}, the pseudo \textbf{Answerer}, the passage \textbf{Summarizer}, and the \textbf{Reranker}. This structured approach enables \ours to (1) more accurately interpret queries, (2) utilize the vast pre-trained knowledge of LLMs, (3) distill passages into concise representations, and (4) comprehensively evaluate passage relevance, resulting in significantly improved reranking outcomes. Our experimental results demonstrate that \ours surpasses state-of-the-art methods on well-established IR benchmarks, including TREC-DL, BEIR, and NovelEval. Notably, on the less contaminated NovelEval dataset, \ours achieves an improvement of over 5 points in NDCG@10 than RankGPT-4. Furthermore, we conduct extensive studies to examine the individual contributions of each role in \ours. 
\end{abstract}



\maketitle

\section{Introduction}

The integration of large language models (LLMs)~\citep{brown2020language, touvron2023llama, lewis2020bart} into Information Retrieval (IR) systems has revolutionized user interactions with information and knowledge~\citep{hou2024large, fan2023recommender, xi2023towards}. LLMs enhance the IR process, including query rewriting and retrieval, through advanced linguistic understanding, semantic representation, context management, and encyclopedic knowledge~\citep{wang2023query2doc, sachan2022improving, qin2023large}.

The application of LLMs to zero-shot text ranking has seen increasing interest. Based on the type of instruction employed, Ranking strategies utilizing LLMs can be categorized into Pointwise~\citep{sachan2022improving, liang2022holistic}, Pairwise~\citep{qin2023large, sun2023instruction}, and Listwise methods~\citep{sun2023chatgpt, pradeep2023rankzephyr}. Among existing methods, listwise approaches achieve superior performance by enabling concurrent relevance judgment across a list of passages~\citep{sun2023chatgpt, pradeep2023rankzephyr}.

\begin{figure*}
    \centering
    \includegraphics[width=0.98\linewidth]{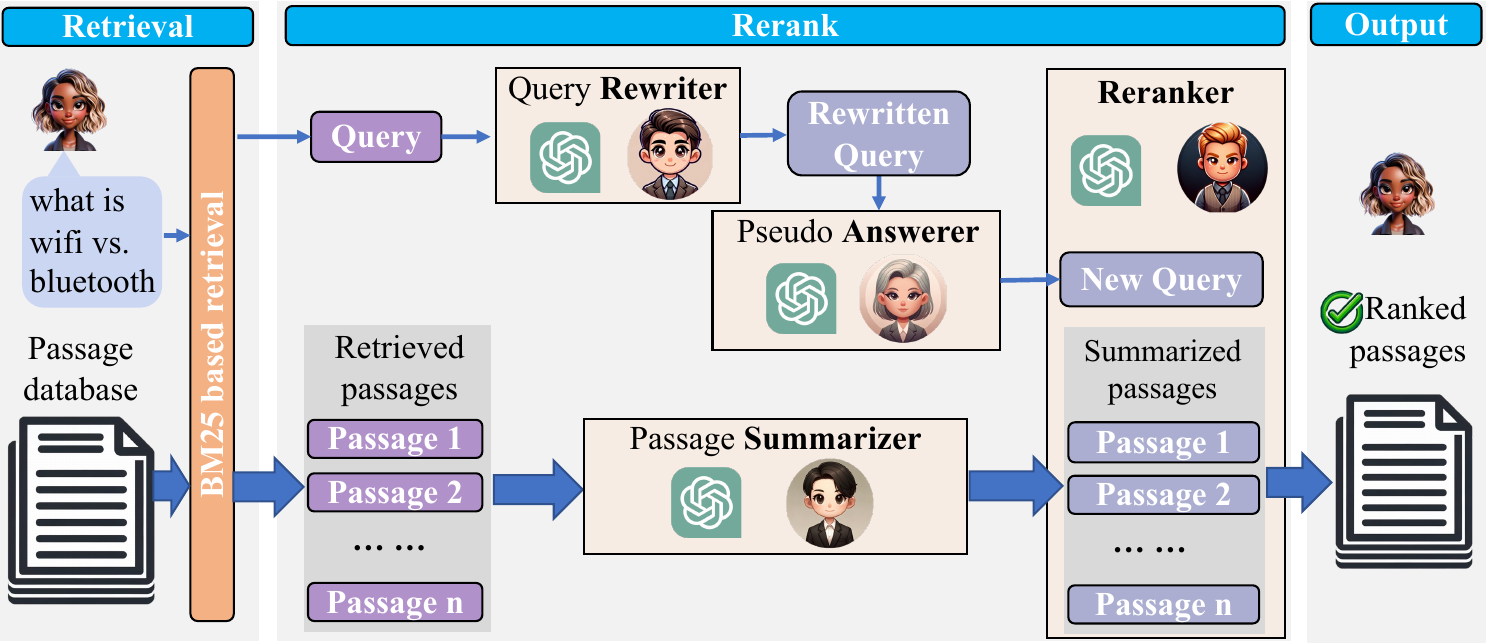}
    \caption{Overview of \ours. \ours is composed of four well-defined expert roles: \textbf{Rewriter}, \textbf{Answerer}, \textbf{Summarizer}, and \textbf{Reranker}, each designed to address specific issues in passage reranking. These roles work sequentially to handle the ranking task.
    }
    \label{figure_method}
\end{figure*}

Although LLMs exhibit strong semantic understanding abilities, retrieval performance can still be compromised by queries that are short, ambiguous, or lack context~\citep{wang2023query2doc, nogueira2019document}. Moreover, the listwise reranking approach, which involves lengthy contexts, faces the challenge of decreased LLM instruction following and reasoning capabilities as context length increases~\citep{sun2023chatgpt, levy2024same, bai2023longbench,jin2025two}, potentially affecting reranking performance. Motivated by the notion that structured workflows enhance task execution consistency and accuracy~\citep{wooldridge1998pitfalls, belbin2022team, hong2023metagpt,liu2022distributed,liu2024learning,li2025comprehensive,xu2025graphomni,zhang2025near}, we introduce \ours—a multi-role workflow utilizing LLMs for reranking. As illustrated in Figure~\ref{figure_method}, \ours applies enriched queries with LLMs' prior knowledge and summarized information from passages to overcome existing challenges in LLM-based reranking systems.

We assess the performance of \ours across a broad range of datasets, including TREC-DL~\citep{craswell2020overview}, BEIR~\citep{thakur2021beir}, and NovelEval~\citep{sun2023chatgpt}. Our empirical findings consistently demonstrate \ours's superior performance. Notably, \ours outpaces current state-of-the-art (SoTA) methods, achieving higher scores than RankGPT~\citep{sun2023chatgpt} on four BEIR datasets—Covid, NFCorpus, SciFact, and Robust04—by an average of $2.5\%$ in nDCG@10, and surpassing RankZephyr~\citep{pradeep2023rankzephyr} on NovelEval by $5\%$ in nDCG@\{1, 5, 10\}.

In a nutshell, our contributions can be highlighted from four perspectives: 
\begin{enumerate}
    \item We introduce a unique multi-role reranking workflow, denoted as \ours, which is grounded in LLMs and employs well-defined role specializations. This workflow exhibits remarkable adaptability, allowing for the dynamic alteration of roles to enhance reranking efficacy.
    \item Our approach innovatively addresses the challenges of semantic ambiguity and context length constraints in listwise zero-shot reranking by incorporating query rewriting and passage summarization to augment clarity.
    \item We carry out extensive experiments on a variety of datasets, evidencing that \ours consistently surpasses SoTA methods~\citep{sun2023chatgpt, pradeep2023rankzephyr}.
    \item We thoroughly examine the impact of individual roles within \ours, providing valuable insights for further investigation.
\end{enumerate}

\begin{table*}[ht]
\centering
\caption{Examples of queries and passages from RankGPT and \ours using GPT-4 on TREC-DL19. \textcolor{red}{red} highlights ambiguous information, \textcolor{yellow}{yellow} indicates vague or infrequently used expressions, and \textcolor{blue}{blue} marks redundant information. \ours demonstrates better ranking performance compared to RankGPT on the same query by leveraging more accurately interpreted queries and clearer, more concise passages.}
\label{table_qualitative_analysis}
\begin{tabular}{|p{1.6cm}|p{6.5cm}|p{6.5cm}|}
\toprule
\textbf{} & \textbf{RankGPT-4} & \textbf{\ours} \\ \midrule
\textbf{Query} & 
\small \textcolor{red}{what} is wifi \textcolor{yellow}{vs} Bluetooth & 
\small Provide a comparative analysis of Wi-Fi and Bluetooth, detailing their differences and similarities. \\ \midrule
\textbf{Passage} & 
\small NFC, or Near Field Communication, is essentially just another transceiver inside your phone or tablet. NFC can let you share information between phones, read information from a sticker or sign, or even buy good and services as easily as you would with a credit card. What differentiates NFC from your Bluetooth, WiFi, or cellular radios is primarily the distance it can cover, and the power it requires to run. NFC runs on a ridiculously little amount of power, but requires a fairly large antenna to work. \textcolor{red}{FC} \textcolor{blue}{can let you share information between phones, read information from a sticker or sign, or even buy good and services as easily as you would with a credit card. What differentiates NFC from your Bluetooth, WiFi, or cellular radios is primarily the distance it can cover, and the power it requires to run.} & 
\small NFC, or Near Field Communication, is a transceiver in devices like phones or tablets that allows for sharing information, reading data from stickers or signs, and making purchases similar to a credit card. It differs from Bluetooth, WiFi, or cellular radios due to its limited coverage distance and low power requirement, despite needing a large antenna to function. \\ \midrule
\textbf{nDCG@10} & 
62.13 & 
\textbf{66.80} \\ 
\bottomrule
\end{tabular}
\end{table*}

\section{Related Works}
\subsection{LLMs for Information Retrieval}
Text retrieval is a key component in a multitude of knowledge-driven Natural Language Processing (NLP) applications \citep{jin2024learning,jin2024apeer,wu2024cg,Weng2024comprehensive,weng2024leveraging,he2024t,he2026sakestructuredagenticknowledge,zuo2025intelligent,wu2024supporting}. In practice, this task is approached with a multi-stage ranking pipeline, typically consisting of an initial, cost-effective retriever followed by a more sophisticated reranker to refine the results~\citep{ma2023zero, craswell2020overview, nogueira2019multi}. Large language models (LLMs) have shown remarkable efficacy in information retrieval tasks~\citep{zhu2023large, sun2023chatgpt, pradeep2023rankzephyr}. Supervised reranking methods~\citep{nogueira2020document, zhuang2023rankt5, pradeep2023rankzephyr} have traditionally relied on fine-tuning transformer-based models with copious training data, such as the MS MARCO v1 passage ranking dataset~\citep{bajaj2016ms}. However, recent explorations involve LLMs in zero-shot unsupervised reranking. Pointwise approaches evaluate passage relevance individually~\citep{sachan2022improving, liang2022holistic}, whereas pairwise strategies compare two documents' relevancies for a given query~\citep{qin2023large, sun2023instruction}. Listwise methods, which directly reorder document relevance collectively, have achieved state-of-the-art performance~\citep{sun2023chatgpt, ma2023zero}. This study introduces a novel multi-role reranking framework, \ours, which significantly enhances listwise reranking performance.

\subsection{Query Rewriting}
Original queries in traditional IR systems are often short or ambiguous, leading to vocabulary mismatch issues. Classic query rewriting techniques refine the original query iteratively by analyzing top-retrieved documents~\citep{abdul2004umass, metzler2005markov, zhai2001model, metzler2007latent,nie2024code}. These methods, however, largely depend on term frequency statistics and may not grasp the true query intent. LLMs, with their advanced linguistic capabilities, support the generation of query rewrites that more accurately reflect the complex and varied information needs of users~\citep{mao2023large, gao2023precise, jagerman2023query, ma2023query,wang2024tool,yu2024credit}. HyDE~\citep{gao2023precise} utilizes dense retrievers to generate pseudo-documents, while Query2doc~\citep{wang2023query2doc} and InPars~\citep{bonifacio2022inpars}, along with Promptagator~\citep{dai2022promptagator}, harness LLMs for producing synthetic queries through zero-shot or few-shot prompting.

\subsection{Prompt Engineer}
Prompt engineering is a critical technique for efficiently tailoring models to specific downstream tasks without fine-tuning~\citep{liu2023pre, brown2020language, zhou2022conditional,jin2023visual,zhou2024adapi,zhang2023online,zhang2022stochastic,zhao2024a}. The chain-of-thought (CoT) prompting method was introduced to encourage LLMs to generate intermediate reasoning steps before reaching a final answer~\citep{kojima2022large, wei2022chain}. In-context learning (ICL) leverages a few examples within the input to guide LLMs towards the intended task~\citep{radford2019language, liu2022few}. Expert prompting~\citep{xu2023expertprompting} designs prompts that emulate an expert's reasoning, tailored to the input query's context. Multi-persona prompting~\citep{du2023improving} employs a range of `personas' to tackle specific tasks. In \ours, we engage LLMs with various roles outlined in a standard operating procedure (SOP) for retrieval, yielding empirically validated improvements in reranking.

\section{\ours Framework Using LLMs}
Section~\ref{method_roles} presents the role specialization and overall procedure in~\ours. Sections~\ref{method_rewiter}, \ref{method_answerer}, \ref{method_summarizer}, and \ref{method_reranker} elucidate the details for each role.

\subsection{Role Specialization and Overall Procedure}\label{method_roles}

Unambiguous role specialization facilitates the decomposition of intricate work into smaller, distinct tasks. In \ours, we divide the reranking task into four parts, assigning four agents with specific skills and expertise for each subtask to degrade the difficulty of the original ranking task. Defining the LLMs' roles and operational abilities establishes a workflow, allowing the LLMs to work sequentially.

As illustrated in Figure~\ref{figure_method} and Table~\ref{table_qualitative_analysis}, the Rewriter, proficient in rephrasing, reformulates the user's query into a clearer and more interpretable version. This refined query is then passed to the Answerer, which generates a standard response. The rewritten query is subsequently concatenated with the generated answer, forming a new query that provides more comprehensive information about the original query. During the ranking process, the passage Summarizer produces concise summaries for each candidate passage, effectively removing redundancies and grammatical errors while preserving only the essential information from the original passage. Finally, the Reranker receives the clarified and enriched query, along with the refined summarized candidate passages, and generates a more accurate, relevance-based reranking result.

\subsection{Rewriter}\label{method_rewiter}
Original queries often exhibit brevity or ambiguity. As shown in Table \ref{table_qualitative_analysis}, a query in TREC-DL19~\citep{craswell2020overview} is `what is wifi vs bluetooth', where the desired passage should emphasize the distinctions and similarities between `wifi' and `bluetooth'. Nevertheless, most pertinent documents in the search results utilize the term `comparative' instead of `vs'. Existing query rewriting techniques employ document corpora to supply domain-specific knowledge for subject areas~\citep{gao2023precise, ma2023query}. These approaches concentrate on examining query rewriting in the initial retrieval stage, while its application in the subsequent reranking stage remains underexplored. In~\ours, we concentrate on harnessing LLMs' language abilities for query rewriting in passage reranking, employing role specialization without reliance on any corpora base.

Let $q$ be a query from the query distribution $\mathcal{Q}$. We specialize the LLM as an expert in refining user queries to enhance their suitability for ranking tasks. Next, we use a zero-shot prompt $c_{rew}$, devoid of any corpus or specific domain knowledge, for query rewriting to obtain a clearer and well-interpreted query $q_{rew}$. $f$ represents the LLM.
\begin{equation}
    q_{rew} = f(q; c_{rew})
\end{equation}

The specialization and prompting for Rewriter can be found in Appendix~\ref{appendix_prompt_rewriter}. Generally, the rewritten query $q_{rew}$ conveys more precise information and avoids ambiguous expressions.

\subsection{Answerer}\label{method_answerer}
Query expansion enhances retrieval systems by enriching query representation with additional terms, expressing identical concepts or information needs, and improving lexical or semantic alignment with corpus documents~\citep{datta2008image, huang2009analyzing}. Early research on query expansion focused on using lexical knowledge bases~\citep{robertson1995okapi} or Pseudo-Relevance Feedback (PRF)~\citep{borgeaud2022improving}. \citet{wang2023query2doc} propose expanding the original query during the previous sparse or dense retrieval stage. Our Answerer adopts a similar approach to \citet{wang2023query2doc}, emphasizing reranking performance enhancement without any knowledge base, necessitating more refined relevance judgment.

We specialize the LLM as an assistant adept at providing detailed and pertinent responses to user queries. Through carefully crafted prompts, the Answerer generates a pseudo-passage $P_{gen}$ that addresses the query, serving as a `standard answer' for the given query $q_{rew}$:
\begin{equation}
    P_{gen} = f (q_{rew}; c_{gen})
\end{equation}
where $c_{gen}$ represents the zero-shot prompt for `standard answer' generation.

Owing to the LLMs' proficient language ability and world knowledge, the generated passage offers abundant information about the given query. The complete specialization for our Answerer is illustrated in Appendix~\ref{appendix_prompt_answerer}.

We then define a new query $q_{new}$ as follows:
\begin{equation}
    q_{new} = \mathrm{Concat}(q_{rew} * m, P_{gen})
\end{equation}
where $*$ denotes string repetition, $m$ represents the number of repetitions, and $\mathrm{Concat}$ is the string concatenation operator. Our experiments demonstrate that repeating the query an appropriate number of times strengthens the query's `weights', leading to improved passage reranking performance.

\subsection{Summarizer}\label{method_summarizer}
As shown in Table \ref{table_qualitative_analysis}, candidate passages $\mathcal{P} = \{P_1, P_2, ..., P_n\}$ from the preceding retrieval stage are often lengthy, and their contained information may be vague, ambiguous, or redundant, complicating concise relevance judgments. To tackle this issue, we suggest summarizing candidate passages before utilizing them for reranking. These summaries effectively preserve essential information without redundancy and are typically much shorter than the original passages, facilitating improved relevance judgment.

To generate enhanced summaries of the original passages, we devise a Summarizer skilled in condensing passages for better information retrieval. For each candidate passage $P\in\mathcal{P}$, we obtain a summarized passage $\hat{P}$ as follows:
\begin{equation}
    \hat{P} = f(P; c_{sum})
\end{equation}
where $c_{sum}$ represents the zero-shot prompt for passage summarization. The complete prompt can be found in Appendix~\ref{appendix_prompt_summarizer}.

Following the summarization process, we acquire a list of summarized candidate passages $\hat{\mathcal{P}}=\{\hat{P}_1, \hat{P}_2, ..., \hat{P}_n\}$.

\subsection{Reranker}\label{method_reranker}
\citet{sun2023chatgpt} propose listwise permutation generation to directly output a ranked list given a set of candidate passages. However, the listwise approach necessitates a considerable number of tokens, potentially degrading instruction following and reasoning abilities~\citep{levy2024same, bai2023longbench, sun2023instruction} and negatively affecting ranking performance. In~\ours, we mitigate this limitation by employing summarized passages for listwise reranking, which are shorter and clearer than the original ones. To enable more precise relevance judgments and further enhance ranking performance through increased reasoning, we adopt a distinct prompting strategy from~\citet{sun2023chatgpt}, encompassing three aspects:

\begin{itemize}
    \item \textbf{Relevance Standard.} To facilitate more accurate relevance judgments, we instruct the LLM to adhere to a detailed relevance standard, as proposed by~\citet{craswell2020overview}. For instance, a passage is deemed perfectly relevant if it is dedicated to the query and contains the answer, whereas it is considered irrelevant if it bears no relation to the query.
    \item \textbf{CoT.} CoT prompting can elicit the reasoning ability of LLMs~\citep{wei2022chain}. Prior studies have applied CoT in query rewriting~\citep{jagerman2023query, alaofi2023can}, while the impact of CoT in passage reranking remains underexplored. In~\ours, we prompt the LLM to rank passages thoughtfully and systematically, enabling more reasoning in the relevance judgment process.
    \item \textbf{Format Requirement.} After employing the relevance standard and CoT, the LLM's output format becomes more diverse (e.g., containing rationales). To minimize malformed outputs, we instruct the LLM to conform to the ranking format and ensure that no passages are omitted or repeated in the ranking results.
\end{itemize}

In~\ours, we specialize the LLM as an adept intelligent assistant for ranking passages based on query relevance. Our carefully designed prompt strategy enables the Reranker to generate superior ranking results.

Assuming there are $n$ summarized candidate passages $\hat{\mathcal{P}}=\{\hat{P}_1, \hat{P}_2, ..., \hat{P}_n\}$ from the Summarizer, we rerank these passages in a back-to-first order from $\hat{P_n}$ to $\hat{P}_1$ using a sliding window of length $w$. Initially, we attain a rank list $l_1$ by prompting the LLM to rank the passages $[\hat{P}_{n-w+1}, \dots, \hat{P}_{n-1}, \hat{P}_n]$ according to their relevance to $q_{new}$:
\begin{equation}
    l_1 = f(q_{new}, [\hat{P}_{n-w+1}, ..., \hat{P}_{n-1}, \hat{P}_n]; c_{list})
    \label{equation_reranker}
\end{equation}
where $c_{list}$ represents our zero-shot listwise prompt, employing the relevance standard, CoT, and format requirement.

Subsequently, we reorder the passage order based on the rank list $l_1$ to obtain a ranked passage set, still denoted as $\hat{\mathcal{P}}=\{\hat{P}_1, \hat{P}_2, ..., \hat{P}_n\}$ for simplicity. We then slide the window in steps of length $s$ and rank the passages $[\hat{P}_{n-s-w+1}, ..., \hat{P}_{n-s-1}, \hat{P}_{n-s}]$ following equation~\ref{equation_reranker}, resulting in a rank list $l_2$. We reorder the passage order in $\hat{\mathcal{P}}$ based on $l_2$. This procedure repeats until all passages are ranked.

\begin{table*}[ht]
\centering
\caption{\textbf{Results (nDCG@$\{$1,5,10$\}$) on TREC.} The performance of eight reranking methods on TREC-DL19 and TREC-DL20. All the unsupervised methods use zero-shot prompts. The best performances are marked in bold.}
\label{table_trec}
\resizebox{0.95\textwidth}{!}{
\begin{tabular}{l c | ccc | ccc }
\toprule
 \multirow{2}{*}{\textbf{Method}} & \multirow{2}{*}{\textbf{LLM}} & \multicolumn{3}{c}{\textbf{TREC-DL19}} & \multicolumn{3}{c}{\textbf{TREC-DL20}} \\
  & & nDCG@1 & nDCG@5 & nDCG@10 & nDCG@1 & nDCG@5 & nDCG@10 \\
\midrule
BM25 & - & 54.26 & 52.78 & 50.58 & 57.72 & 50.67 & 47.96 \\ 
\midrule
\textbf{Supervised} \\ 
\midrule
monoT5 & T5 (3B) & 79.04 & 73.74 & 71.83 & 80.25 & 72.32 & 68.89 \\ 
RankT5 & T5 (3B) & 77.38 & 73.94 & 71.22 & 80.86 & 72.99 & 69.49 \\ 
RankZephyr & Zephyr (7B) & - & - & 74.20 & - & - & 70.86 \\ 
\midrule
\textbf{Unsupervised} \\ 
\midrule
RG & FLAN-UL2 (20B) & 70.93 & 66.81 & 64.61 & 75.62 & 66.85 & 65.39 \\ 
PRP & FLAN-UL2 (20B) & 78.29 & 75.49 & 72.65 & \textbf{85.80} & 75.35 & 70.46 \\ 
RankGPT-4 & GPT-4 & 80.62 & 77.83 & 74.89 & 79.73 & 73.15 & 70.14   \\ 
\midrule
\ours & GPT-4 & \textbf{83.33} & \textbf{79.44} & \textbf{76.65}   & 82.41 & \textbf{75.68} & \textbf{71.80}  \\ 
\bottomrule
\end{tabular}}
\end{table*}

\section{Experiments}
To assess the efficacy of~\ours, we perform extensive experiments to: (1) showcase the superior performance of~\ours across various benchmarks; (2) examine the qualitative and quantitative influence of each component in our approach; and (3) present the results of \ours on diverse models, along with findings from ablation study.

\subsection{Models and Benchmarks}\label{models_and_benchmarks}
We choose GPT-4~\citep{achiam2023gpt} as our primary model, employing the Azure API, which features a GPT-4-0613 version. Our experiments are evaluated on three benchmark datasets, including TREC-DL~\citep{craswell2020overview}, BEIR~\citep{thakur2021beir}, and NovelEval~\citep{sun2023chatgpt}. 

\textbf{TREC} is a widely adopted benchmark in IR research. We use the test sets from the TREC-DL19 and TREC-DL20, which employed the MS MARCO v1 passages.

\textbf{BEIR} encompasses diverse retrieval tasks and domains. We select the test sets of eight tasks in BEIR to evaluate our approach: \begin{enumerate*}[label=(\roman*)]
\item \emph{Covid}: Retrieves scientific articles related to COVID-19.
\item \emph{NFCorpus}: A biomedical information retrieval dataset.
\item \emph{SciFact}: Retrieves evidence for claims verification.
\item \emph{Robust04}: Assesses challenging topics.
\item \emph{Touche}: An argument retrieval dataset.
\item \emph{DBPedia}: Retrieves entities from the DBpedia corpus.
\item \emph{Signal}: Retrieves relevant tweets for a given news title.
\item \emph{News}: Retrieves relevant news articles for headlines.
\end{enumerate*}

\textbf{NovelEval} features queries not learned by GPT-4-0613~\citep{sun2023chatgpt}. Questions in current benchmarks (e.g., TREC-DL) are typically collected years ago, raising concerns that existing LLMs may already possess knowledge of these questions~\citep{yu2023generate}. Moreover, since many LLMs do not disclose information about their training data, there is a potential risk of contamination in existing benchmark test sets~\citep{achiam2023gpt}. To mitigate these concerns, we evaluate~\ours on NovelEval-2306.

\begin{table*}[htbp]
  \centering
      \caption{\textbf{Results (nDCG@10) on BEIR.} The performance of six reranking methods on eight BEIR datasets. RankT5 reranks the top 1000 passages returned by BM25 while other methods rerank top 100 passages by BM25.}
    \label{table_beir}
    \resizebox{0.95\textwidth}{!}{
    \begin{tabular}{l | cccccccc | c }
      \toprule
       \textbf{Method} & \textbf{COVID} & \textbf{NFCorpus} & \textbf{SciFact} & \textbf{Robust04} & \textbf{Touche} & \textbf{DBpedia} & \textbf{Signal} & \textbf{News} & \textbf{Avg} \\
      \midrule
      BM25 & 59.47 & 30.75 & 67.89 & 40.70 & \textbf{44.22} & 31.80 & 33.05 & 39.52 & 43.42 \\ 
      \midrule
      monoT5 & 80.71 & 38.97 & 76.57 & 56.71 & 32.41 & 44.45 & 32.55 & 48.49 & 51.36 \\ 
      RankT5 & 80.71 & 38.10 & 74.99 & - & 44.01 & 44.22 & 32.00 & - & - \\ 
      RankZephyr & 83.78 & - & - & - & - & - & - & 51.84 & - \\
      \midrule
      RankGPT-4 & 83.98 & 38.83 & 75.61 & 59.74 & 40.72 & 47.12 & 33.90 & 52.82 & 54.09  \\ 
      \midrule
      \textbf{\ours} & \textbf{85.77} & \textbf{39.74} & \textbf{77.73} & \textbf{64.88} & 42.08 & \textbf{47.43} & \textbf{34.54} & \textbf{52.97} & \textbf{55.64}  \\ 
      \bottomrule
    \end{tabular}}
\end{table*}

\begin{table}[ht]
\centering
    \caption{\textbf{Results (nDCG@$\{$1,5,10$\}$) on NovelEval.}}
    \label{table_noveleval}
  \resizebox{0.45\textwidth}{!}{
    \begin{tabular}{l | ccc }
      \toprule
      \textbf{Method} & nDCG@1 & nDCG@5 & nDCG@10 \\
      \midrule
      BM25 & 33.33 & 45.96 & 55.77 \\ 
      \midrule
      monoT5 & 83.33 & 78.38 & 84.62 \\
      RankZephyr & 92.86 & 86.15 & 89.34 \\
      \midrule
      RankGPT-4 & 92.86 & 86.10 & 89.18  \\ 
      \midrule
      \textbf{\ours} & \textbf{97.62} & \textbf{91.79} & \textbf{94.21} \\ 
      \bottomrule
    \end{tabular}}
  \end{table}

\begin{table*}[ht]
\centering
\caption{nDCG@$\{$1,5,10$\}$ of utilizing Rewriter. “RankGPT-4 + Rewriter” incorporates our Rewriter in RankGPT-4.}
\label{table_rq}
\footnotesize 
\setlength\tabcolsep{10pt} 
\resizebox{0.9\textwidth}{!}{
\begin{tabular}{l | ccc }
\toprule
\textbf{Method} & \textbf{TREC-DL20} & \textbf{COVID} & \textbf{NovelEval} \\
\midrule
RankGPT-4 & 79.73/73.15/70.14 & 88.25/86.67/83.98 & 92.86/86.10/89.18  \\ 
\midrule
RankGPT-4 + Rewriter & 80.56/74.06/70.10 & 91.33/87.06/84.50 & 91.27/86.77/91.50  \\ 
\midrule
\ours & \textbf{82.41}/\textbf{75.68}/\textbf{71.80} & \textbf{96.00}/\textbf{89.48}/\textbf{85.77} & \textbf{97.62}/\textbf{91.79}/\textbf{94.21} \\ 
\bottomrule
\end{tabular}}
\end{table*}

\subsection{Baselines and Evaluation Metrics}\label{baselines_and_metrics}
\paragraph{Baselines.} We select several representative SoTA passage reranking methods as our baselines: (1) \textit{BM25}~\citep{lin2021pyserini} serves as a fundamental sanity check in reranking, directly using the rank results after the previous retrieval stage; (2) \textit{monoT5}~\citep{nogueira2020document} is a sequence-to-sequence reranker employing T5 (3B) to compute the relevance score with pointwise ranking loss, trained on MS MARCO; (3) \textit{RankT5}~\citep{zhuang2023rankt5} is a reranker utilizing T5 (3B) and listwise ranking loss, trained on MS MARCO; (4) \textit{RG}~\citep{liang2022holistic} is a pointwise reranking approach based on relevance generation using FLAN-UL2 (20B); (5) \textit{PRP}~\citep{qin2023large} is a pairwise reranking approach employing a sliding window strategy with 10 passes, using the FLAN-UL2 (20B) model; (6) \textit{RankZephyr}~\citep{pradeep2023rankzephyr} is a recent reranker leveraging the 7B parameter Zephyr$_{\beta}$ (built on Mistral), distilled from GPT-3.5 and GPT-4 on MS MARCO; and (7) \textit{rankGPT-4}~\citep{sun2023chatgpt} is our most crucial baseline, adopting a listwise reranking strategy with GPT-4.

\paragraph{Implementation and Metrics.} 
All baselines and~\ours rerank the top $100$ passages retrieved by BM25 using pyserini~\citep{lin2021pyserini} unless specified otherwise. We employ normalized Discounted Cumulative Gain (nDCG) at rank cutoffs of $\{1,5,10\}$ (nDCG@$\{1,5,10\}$) to evaluate performance. For~\ours and rankGPT-4, we utilize the Azure API with a context size setting of $8192$, employing the GPT-4 version GPT-4-0613, which differs from the one used in~\citet{sun2023chatgpt}. Additionally, rankGPT-4 employs GPT-4 to rerank the top $30$ passages reranked by GPT-3.5 (which reranks the top $100$ passages by BM25) on BEIR. These differences result in discrepancies between the rankGPT-4 outcomes in our paper and those in~\citet{sun2023chatgpt}. In our Answerer, we set a repeat time $m$ of $3$. In our Reranker, we use a window size $w$ of $20$ and a step size $s$ of $10$, following~\citet{sun2023chatgpt}. We set the temperature to $0$ for the GPT-4 API to reduce randomness. All our results are averaged over $3$ runs.

\begin{table*}[ht]
\centering
\caption{nDCG@$\{$1,5,10$\}$ of utilizing Answerer. “RankGPT-4 + Answerer” integrates our Answerer in RankGPT-4.}
\label{table_pg}
\footnotesize 
\setlength\tabcolsep{10pt}
\resizebox{1\textwidth}{!}{
\begin{tabular}{l | ccc }
\toprule
\textbf{Method} & \textbf{TREC-DL19} & \textbf{COVID} & \textbf{NovelEval} \\
\midrule
RankGPT-4 & 80.62/77.83/74.89 & 88.25/86.67/83.98 & 92.86/86.10/89.18  \\ 
\midrule
RankGPT-4 + Answerer, $m=1$ & 81.79/77.98/74.21 & 87.67/86.68/84.00 & 97.62/90.51/92.38  \\ 
RankGPT-4 + Answerer, $m=3$ & 82.17/78.41/74.70 & 89.67/88.40/85.00 & 97.62/91.63/93.85  \\ 
RankGPT-4 + Answerer, $m=10$ & 80.82/78.09/74.63 & 89.00/87.87/84.14 & 97.62/91.49/93.83  \\ 
\midrule
\ours & \textbf{83.33}/\textbf{79.44}/\textbf{76.65} & \textbf{96.00}/\textbf{89.48}/\textbf{85.77} & \textbf{97.62}/\textbf{91.79}/\textbf{94.21} \\ 
\bottomrule
\end{tabular}}
\end{table*}

\subsection{Main Results}
\paragraph{Results on TREC.}
To demonstrate the superior performance of \ours on TREC datasets, we compare it with seven baselines. The nDCG@$\{1,5,10\}$ results are presented in Table~\ref{table_trec}. We can draw the following positive observations: \ding{182} \ours exhibits superior performance compared to RankGPT-4 on TREC datasets, surpassing RankGPT-4 by an average nDCG@$\{1,5,10\}$ of $\{2.70, 2.07, 1.71\}$ on TREC-DL19 and TREC-DL20. \ding{183} \ours achieves the best performance among both supervised and unsupervised methods in terms of nDCG@5 and nDCG@10. It not only surpasses all supervised methods but also outperforms SoTA zero-shot unsupervised methods, including the pointwise method RG, the pairwise method PRP, and the listwise method RankGPT-4. This further indicates the effectiveness of our reranking framework.

\paragraph{Resutls on BEIR.}
We further evaluate the performance of~\ours on eight BEIR datasets, which contain more queries and heterogeneous topics than TREC-DL19 and TREC-DL20. The results are displayed in Table~\ref{table_beir}, from which we can observe that: \ding{182}~\ours outperforms the baselines on BEIR, achieving the best nDCG@10 across all baselines. \ding{183}~\ours is robust to diverse topics and queries. It surpasses RankGPT-4 by an average of $\mathbf{1.55}$ on eight BEIR datasets, which contain more queries than the TREC datasets. Notably,~\ours achieves a $\mathbf{5.14}$ nDCG@10 improvement over RankGPT-4 on Robust04, which consists of 249 queries and diverse topics in news.

\paragraph{Results on NovelEval.}
To address the concern of data contamination in LLM reranking, we further evaluate the performance of five reranking methods on NovelEval-2306. The results are shown in Table~\ref{table_noveleval}, from which we can observe that: \ding{182}~\ours maintains significant better reranking performance on unlearned datasets, achieving the best performance among all reranking methods. \ding{183}~\ours exhibits a substantial performance enhancement compared to RankGPT-4 and RankZephyr, with an improvement of $\mathbf{5}$ points in terms of nDCG@1, nDCG@5, and nDCG@10.

\subsection{The Impact of Each Component}\label{section_impact_of_component}
\paragraph{Rewriter.}\label{appendix_results_rewriter}
We investigate the effect of the Rewriter in our multi-role reranking workflow. We incorporate the Rewriter into RankGPT-4, which rewrites the original queries in the benchmarks and utilizes the rewritten queries for listwise passage reranking. As shown in Table~\ref{table_rq}, the Rewriter demonstrates a capability to slightly enhance ranking performance by using well-interpreted queries. On TREC-DL20, COVID, and NovelEval, it achieves improvements in the majority of instances and yields an average nDCG@$\{1,5,10\}$ improvement of $\{0.76, 0.66, 0.93\}$.

\paragraph{Answerer.}\label{appendix_results_answer}
We investigate the effects of our Answerer in reranking by integrating it with RankGPT-4. For an original query $q$ from the benchmarks, we generate a passage $P$ answering the query using the Answerer. We then form a new query by repeating $q$ for $m$ times and concatenating the repeated queries with $P$, which is utilized in listwise reranking. The empirical results in Table~\ref{table_pg} yield several positive observations: \ding{182} Our Answerer can enhance the overall ranking performance. For $m \in \{1, 3, 10\}$, incorporating the Answerer leads to performance gains in most cases on TREC-DL19, COVID, and NovelEval. \ding{183} A moderate value of $m$ in the Answerer results in the best performance gains. As shown in Table~\ref{table_pg}, $m=3$ consistently outperforms $m=1$ and $m=10$. \ding{184} The Answerer is capable of generating valuable feedback, even on datasets unlearned by GPT-4. On NovelEval, it achieves performance comparable to \ours. These observations demonstrate that the answers generated by the Answerer contain rich information that can improve semantic-level matching and relevance judgment in reranking.

\paragraph{Summarizer.}\label{appendix_results_summarizer}
We evaluate the effects of Summarizer by integrating the Summarizer into RankGPT-4, which uses the original query from benchmarks but replaces the candidate passages retrieved by BM25 with the summarized passages generated by the Summarizer. The results in Table~\ref{table_sum} reveal that: \ding{182} The Summarizer is capable of enhancing reranking performance, achieving consistent performance gains in terms of nDCG@$\{1,5,10\}$ on TREC-DL20, COVID, and Robust04. \ding{183} The Summarizer particularly improves the performance of nDCG@10, achieving an average nDCG@10 improvement of $2.14$ on three datasets and comparable nDCG@{10} with \ours. Notably, it attains an nDCG@10 improvement of $\mathbf{3.95}$ on Robust04. The findings indicate that substituting vague, ambiguous, and redundant passages with concise, clear, and well-structured ones improves reranking performance.

\begin{table*}[ht]
\centering
\caption{nDCG@$\{$1,5,10$\}$ of utilizing the Summarizer. “RankGPT-4 + Summarizer” integrates our Summarizer into RankGPT-4.}
\label{table_sum}
\footnotesize 
\setlength\tabcolsep{10pt}
\resizebox{0.9\textwidth}{!}{
\begin{tabular}{l | ccc }
\toprule
\textbf{Method} & \textbf{TREC-DL20} & \textbf{COVID} & \textbf{Robust04} \\
\midrule
RankGPT-4 & 79.73/73.15/70.14 & 88.25/86.67/83.98 & 75.30/66.07/59.74  \\ 
\midrule
RankGPT-4 + Summarizer & 81.07/74.46/71.34 & 92.00/87.06/85.26 & 78.58/70.44/63.69  \\ 
\midrule
\ours & \textbf{82.41}/\textbf{75.68}/\textbf{71.80} & \textbf{96.00}/\textbf{89.48}/\textbf{85.77} & \textbf{80.92}/\textbf{70.99}/\textbf{64.88} \\ 
\bottomrule
\end{tabular}}
\end{table*}

\paragraph{Reranker.}\label{appendix_results_reranker}
We conduct additional experiments to investigate the effect of different prompt designs in the Reranker. As indicated in Section~\ref{method_reranker}, our prompting strategy comprises three aspects: Relevance Standard, CoT, and Format Requirement. We add each type of prompt to RankGPT-4 separately, resulting in `RankGPT-4 w. Relevance Standard', `RankGPT-4 w. CoT', and `RankGPT-4 w. Format Requirement'. The experimental results are presented in Table~\ref{table_prompt}. 

\textit{Relevance Standard.} We prompt RankGPT-4 to follow a detailed four-level relevance standard: Perfectly relevant, Highly Relevant, Relevant, and Irrelevant, as described by \citet{craswell2020overview}. The results in Table~\ref{table_prompt} show that this prompting strategy yields consistent performance gains across all three datasets, indicating that detailed relevance standards can enhance relevance judgments.

\textit{CoT.} We utilize zero-shot CoT prompts in RankGPT-4, instructing it to think thoughtfully and systematically during ranking. We observe that CoT leads to significant performance gains by enabling more reasoning in relevance judgments. `RankGPT-4 w. CoT' achieves consistent performance improvements, with an average gain of $\{1.99, 2.15, 1.97\}$ in terms of nDCG@$\{1,5,10\}$ compared to RankGPT-4 on the three datasets.

\textit{Format Requirement.} We incorporate format requirement prompts to instruct RankGPT-4 to adhere to the specific rank format and ensure no repeated or missing passages in the rank list, facilitating a more convenient extraction of the final rank results. As shown in Table~\ref{table_prompt}, the format requirement instructions marginally improve reranking performance in most cases, yielding an average gain of $\{1.26, 0.69, 1.05\}$ in terms of nDCG@$\{1,5,10\}$ on TREC-DL20, COVID, and NovelEval.

\begin{table*}[ht]
\centering
\caption{nDCG@$\{$1,5,10$\}$ of utilizing different prompting strategies in 
the Reranker. “RankGPT-4 w. Relevance Standard”, “RankGPT-4 w. CoT”, and “RankGPT-4 w. Format Requirement” indicate incorporating our Relevance Standard, CoT prompting, and Format Requirement into RankGPT-4, respectively.}
\label{table_prompt}
\footnotesize 
\setlength\tabcolsep{10pt}
\resizebox{0.9\textwidth}{!}{
\begin{tabular}{l | ccc }
\toprule
\textbf{Method} & \textbf{TREC-DL20} & \textbf{COVID} & \textbf{NovelEval} \\
\midrule
RankGPT-4 & 79.73/73.15/70.14 & 88.25/86.67/83.98 & 92.86/86.10/89.18  \\ 
\midrule
RankGPT-4 w. Relevance Standard & 80.45/73.94/70.62 & 91.67/88.04/85.38 & 92.89/89.57/92.49  \\ 
RankGPT-4 w. CoT & 80.25/74.13/70.42 & 93.33/88.51/85.51 & 95.24/89.74/93.28  \\ 
RankGPT-4 w. Format Requirement & 79.63/73.68/70.38 & 91.33/87.11/84.70 & 93.65/87.19/91.24  \\ 
\midrule
\ours & \textbf{82.41}/\textbf{75.68}/\textbf{71.80} & \textbf{96.00}/\textbf{89.48}/\textbf{85.77} & \textbf{97.62}/\textbf{91.79}/\textbf{94.21} \\ 
\bottomrule
\end{tabular}}
\end{table*}

\begin{table*}[ht]
\centering
\caption{Average token cost, running time, and USD cost per query on TREC-DL.}
\label{table_cost_comparison}
\begin{tabular}{lcccc}
\toprule
\textbf{Dataset} & \textbf{ \# Input Tokens} & \textbf{\# Output Tokens} & \textbf{Time (s)} & \textbf{USD} \\
\midrule
RankGPT-4 & 19,890 & 732 & 82 & 0.641 \\
\ours & 21,938 & 5,507 & 76 & 0.989 \\
\ours w. Local Passages & 12,027 & 1,500 & 60 & 0.451 \\
\bottomrule
\end{tabular}
\end{table*}

\begin{table}[t]
\centering
    \caption{Results (nDCG@10) on LLaMA-3-8B, LLaMA-3-70B, and GPT-3.5-Turbo-0301.}
    \label{table_results_llama}
  \resizebox{0.5\textwidth}{!}{
    \begin{tabular}{l | c| c c }
      \toprule
      \textbf{Model} & \textbf{Method} & \textbf{TREC-DL19} & \textbf{TREC-DL20} \\
      \midrule
      GPT-3.5 & RankGPT & 65.80 & 62.91 \\ 
      GPT-3.5 & \ours & \textbf{67.87} & \textbf{64.01} \\
      \midrule
      LLaMA-3-8B & RankGPT & 56.07 & 52.81 \\ 
      LLaMA-3-8B & \ours & \textbf{61.85} & \textbf{57.82} \\
      \midrule
      LLaMA-3-70B & RankGPT & 71.45 & 67.42 \\ 
      LLaMA-3-70B & \ours & \textbf{73.34} & \textbf{69.07} \\
      \bottomrule
    \end{tabular}}
\end{table}

\subsection{Additional Investigation}\label{section_additional_investigation}

\paragraph{Results on Diverse Models.} We conduct additional experiments on DL19 and DL20 using LLaMA-3-8B, LLaMA-3-70B~\citep{llama3modelcard}, and GPT-3.5-Turbo-0301~\citep{ChatGPT}. As shown in Table~\ref{table_results_llama}, \ours consistently outperforms RankGPT across all models and datasets. Notably, \ours proves particularly effective on smaller models such as LLaMA-3-8B, achieving more than a 5-point improvement in nDCG@10 on both DL19 and DL20. It is important to highlight that LLaMA-3-8B struggles with following instructions in RankGPT, likely due to the excessive length of the input, which exceeds the model's comprehension capabilities. In contrast, \ours leverages shorter summaries during reranking, making it significantly more effective than RankGPT on smaller models.

\paragraph{Cost Analysis.} We compare the average token cost, runtime, and USD cost per query of \ours with RankGPT using GPT-4-0613. In \ours, we can store the rewritten queries and summarized passages locally for reuse in subsequent reranking processes, thus reducing both runtime and financial costs. We refer to this approach as “\ours w. Local Passage”. The results are presented in Table~\ref{table_cost_comparison}. The key observations are as follows: \ding{182} While \ours incurs higher token costs than RankGPT, it operates faster and achieves an average nDCG@10 improvement of nearly 2\% on the TREC, BEIR, and NovelEval datasets, as shown in Tables~\ref{table_trec}, \ref{table_beir}, and \ref{table_noveleval}. The additional tokens in \ours stem from the inclusion of extra content in the ranking results, such as the analysis process. \ours is faster because it uses significantly shorter passages during reranking compared to RankGPT. \ding{183} The “\ours w. Local Passage” variant effectively reduces token costs and financial expenses, saving 29.64\% in USD and operating 26.83\% faster than RankGPT.

\section{Conclusion}
In this paper, we introduce \ours, a novel multi-role collaborative reranking workflow for LLMs comprising four roles: Rewriter, Answerer, Summarizer, and Reranker. Our extensive empirical results showcase our significant effectiveness, which consistently surpasses SoTA methods on various datasets. Furthermore, we investigate the individual contributions of each role and the impacts of prompt designs, providing valuable insights for future research.

\section{Acknowledgments}

This research is partially funded by research grants to Metaxas from NSF: 2310966, AFOSR 23RT0630, and NIH 2R01HL127661.
\newpage
\bibliographystyle{ACM-Reference-Format}
\bibliography{sample-base}

\appendix

\section{Specilization and Prompt for Rewriter}\label{appendix_prompt_rewriter}
\begin{tcolorbox}[colback=gray!5!white,colframe=gray!75!black,boxsep=2pt,left=2pt,right=2pt,top=2pt,bottom=2pt]
\textbf{system} \\
You are an AI retrieval assistant, skilled at rewriting user queries to enhance their suitability for retrieval tasks and optimizing compatibility with retrieval systems like BM25. \\

\textbf{user} \\
Rewrite the following user query into a clear, specific, and formal request suitable for retrieving relevant information from a list of passages. Keep in mind that your rewritten query will be sent to rerank system, which does relevance search for retrieving documents. \\

\textbf{assistant} \\
Kindly provide the query you would like me to rewrite. \\

\textbf{user} \\
\{query\}
\end{tcolorbox}

\section{Specilization and Prompt for Answer}\label{appendix_prompt_answerer}
\begin{tcolorbox}[colback=gray!5!white,colframe=gray!75!black,boxsep=2pt,left=2pt,right=2pt,top=2pt,bottom=2pt]
\textbf{system} \\
You are an AI retrieval expert, skilled at providing detailed and relevant answers to user queries. \\

\textbf{user} \\
Compose a passage to address the following user query effectively. \\

\textbf{assistant} \\
Please provide the query for which you would like an answer. \\

\textbf{user} \\
\{query\}
\end{tcolorbox}

\section{Specilization and Prompt for Summarizer}\label{appendix_prompt_summarizer}
\begin{tcolorbox}[colback=gray!5!white,colframe=gray!75!black,boxsep=2pt,left=2pt,right=2pt,top=2pt,bottom=2pt]
\textbf{system} \\
You are an AI assistant who is good at summarizing passages the user provides you. \\

\textbf{user} \\
I will provide you a passage. Summarize the passage to make it suit for a passage retrieval task which means the summarized passages can better reflect the information and the relevance to a giving query than the original passage. 

Passage: \{passage\}
\end{tcolorbox}
\end{document}